# From the Termite Mound to the Stars: Meditations on Discussions With Ilya Prigogine

Paul J. Werbos
National Science Foundation[*], Room 675
Arlington, VA 22203

## Introduction

Dr. Lyudmila Kuzmina recently invited me to submit a paper for this special issue in honor of Ilya Prigogine. In response, we submitted some new technical work on relations between classical statistics and quantum dynamics, which certainly addresses one part of the range of topics Prigogine was interested in. Yet, in the end, it does not do full justice to the interests and visions of the man.

Prigogine's career reflected a period of great transition in the understanding of dynamical systems. I cannot claim to know all the stages of the evolution of his thought – but, when I had a chance to discuss these issues with him in 1994, in connection with a workshop where we had back-to-back talks [1], he conveyed some new thoughts about order, life and the universe very different from the world of closed system thermodynamics he grew up with. In some ways, Prigogine started out believing in the Classical Heat Death as the destiny of all life and all systems, but ended up taking a position which is almost the opposite extreme.

Our discussion with Prigogine touched on issues which may be very important to the *future* of hard-core mathematical physics. But in my view, we are not yet ready to make that future tangible and mathematical. We have many prerequisites we must fulfill first. This essay will describe that discussion, and go on to describe my personal meditations about the larger situation.

Meditations of this sort are not a proper part of what Kuhn calls "normal science." They are not part of the usual careful step-by-step accumulation of knowledge – a self-organizing system which has much in common with the termite mound, a paradigm for self-organizing systems now very popular in the robotics community. But in order to make useful, real progress in the long-term we need sometimes to look beyond our present efforts, up to the stars, in order to develop some sense of direction and purpose and long-term possibilities.

## Prigogine on Alternative Views of Time, Entropy and the Universe

The classical view of life, time and the universe may be summarized as follows. First, all possible closed dynamical systems possess an "entropy function." The "entropy" of any state is really just the logarithm of the probability of that state appearing after the universe reaches statistical equilibrium. Entropy is also a measure of disorder. Therefore, any closed dynamical system which possesses partial order must be just a transitional state, on the way to the inevitable maximization of entropy or disorder – the Heat Death. More formally – if an "entropy function" is accurately represented as the sum or integral of local components, each of which depend on a different set of system variables, then there is no possibility of cross-correlation, pattern or order in equilibrium. Patterns like "life" are ruled out, except for meaningless passing rare coincidences like clouds which appear to form the image of a dog, a "dog" that never barks.

For a long time, it was recognized that *open* systems, like the biosphere of earth, are a bit different from closed systems. Yet the "hope" among classical thermodynamicists was that the powerful machinery of local entropy functions could be extended to open systems as well – and thus that open systems could be proven to have many of the same properties as closed systems. In retrospect, this is an example of the age-old problem in science, where "the man who sells hammers sees the whole world as a pile of nails." (There are many other examples, such as the overuse of linear methods on nonlinear systems in engineering, and the construction of "cognitive models" of the brain without a serious representation of the primary role of emotions[2].) Prigogine himself played a major role in trying to develop a local "entropy production" function, which would play the same role for open systems as entropy does for closed systems. Yet it became clear [3] that this really could not be done in a reasonably complete or general way. The first major threat to the classical way of thinking about open systems was the subject of *chemical oscillations*, which Prigogine at first resisted (based on the previous mainstream paradigm), but finally embraced whole-heartedly. (If only other great leaders in science could be so flexible about learning new things and new approaches!) Chemical oscillations firmly prove that open systems really do *not* have to go towards anything like a heat-death in equilibrium, even in infinite asymptotic time.

---

[*] The views herein are those of the author and do not represent NSF in any way.

The Big Bang view of the universe is a major living inheritance from the earlier classical period. Even after the discovery of chemical oscillations, it was still believed that *closed dynamical systems* must inevitably go towards the heat death. Since the universe as a whole was viewed as one large closed system, it was concluded that life and order could only be transition phenomena in the universe as a *whole*. Life could persist indefinitely on earth so long as the sun shines (it was now understood), but the sun itself is destined to burn out, and, with it, all other suns and other life in the universe. Since life could not possibly persist beyond a certain finite time, it was deduced that life could exist at all only because of a great anomalous initial boundary condition, the "Big Bang." The dispute between "Big Bang" and steady-state adherents continues to this day in serious astrophysics, but there are two major logical considerations which give the Big Bang the upper hand even among skeptics: (1) the issue of entropy; and (2) the issue of red shifts.

At the workshop in 1994, Prigogine asserted quite strongly [1] that we should *no longer* consider entropy to be an issue in deciding between the Big Bang and the steady state version. He asserted that there is a kind of spontaneous symmetry breaking of the arrow of time which occurs in all systems which possess the kinds of complexity we now see in our universe, particularly in light of quantum dynamics as we now see it over Fock-Hilbert space. Prigogine's new viewpoint could be interpreted as yet another deep evolution of thought, due in part to his appreciation of chaos theory and of the Santa Fe school of complex systems analysis.

Personally, I would not go *quite* so far as Prigogine on this last point. Yet perhaps the mainstream today has underestimated him, as it underestimated Einstein, both when they were younger and when they are older. Perhaps the issues of life and order in the universe as a whole should be revisited.

**My Personal Assessment of the Big Bang Issue**

First, I must confess that I did argue with Prigogine on his 1994 formulation. By analogy, the Santa Fe people have learned that *not all* complex systems are *complex adaptive systems*. It takes certain very special properties for a complex system to become a truly intelligent system. The mathematics of such systems is best known in highly mathematical research groups almost unknown to those who write popular books about complexity[4]. Even to evolve time-forwards life and order, special properties are required. In my opinion, these special properties require more than what we see in the standard model of physics. If the standard model of physics were the whole story, we would all be subject to the Great Heat Death anyway, in my opinion. John Wheeler has published an account of how the heat death works when gravity, neutrinos and such are all accounted for – and it remains as grim as the most classical accounts.

But in my paper for the workshop [1], I showed how small *time-asymmetries* in chemical reaction equations can result in interesting dynamics, even for the simplest energy-conserving closed stoichiometric systems. So far as I know, this result in simulation was a new result. It was difficult to *get* this result in a very simple system, but it seems clear that greater complexity would make it easier to generate complex emergent behavior.

How could this possibly be relevant to physics? The standard model is totally time-symmetric, but there is a well-established (but poorly understood) class of physical reactions called the *superweak interactions* [5]. These are not a significant factor in understanding what happens here on earth, but who can say what happens on the larger scale of the universe? Small feedback interactions which have little impact on short time-scales can often dominate the overall state of a larger system, when integrated over long periods of time. Billions of years and trillions of light-years might well be enough. In my paper in [1], I actually described a kind of "strawman model" for how such interactions could allow life, order and a macroscopic forwards arrow of time to persist *indefinitely*, in an infinite universe. The biggest problem with that model was that it required the existence of a lot of "dark matter," which seemed implausible to many people at that time. In retrospect, perhaps I should have tried to publish the prediction more widely before the new data came in in support of dark matter. I did cite a paper summarizing the extensive empirical work of Arp supporting a *non-Doppler* explanation for the cosmic red shift [11].

Crudely speaking, one could say that the superweak interactions would play a role in the larger universe analogous to the role of light hitting the planet earth. From a large-scale mechanical viewpoint, it would be hard to believe that something as elusive and weak as light could be of practical importance to big things like the movement of tanks and other massive objects on earth. If we did not have eyes evolved to see the light, we would probably believe that it is highly unlikely that light could be of importance to physics or life on this planet. (H.G. Wells had an interesting story, "The Country of the Blind," which portrays this point very clearly.) Yet we all know how important this weak radiation really is in shaping life and order on earth. Perhaps the superweak interactions are really just a local manifestation of another kind of radiative system, analogous to light, which might even be rightly called a "light of creation," insofar as it would underlie the flow of free energy which allows the creation of new stars and life.

In my view, the combination of Arp's work and the new insights into thermodynamics make it *totally unnecessary to assume a Big Bang*. The assumption of a Big Bang does not really "buy us" anything. Yes, there are lots of very specific pieces of cosmological data which have been closely tailored to fit very complex special-purpose models; however, without a compelling reason to believe there must be such a cosmic "moment of creation" a mere 10-20 billion years ago, Occam's Razor suggests that the whole idea ought to be reconsidered. Yes, I am assuming that there is some unknown physics at work; however, we *do know* that there is unknown physics at work.

Some readers may wonder how this view of a hard-wired arrow of time could be reconciled with my analysis of *time-symmetry* in physics experiments here on earth (as in [6]). In fact, I only assume strict effective microscopic symmetry for experiments we can do today within the solar system. We do not need to account for microscopic time-asymmetry when we try to clean up the standard model of physics. My view of quantum measurement is totally in agreement with the work of Huw Price, who has argued that we should try to build devices to *detect* regions of time in which the arrow of time flows backwards[7]. According to conventional views of quantum measurement (even in the Aharonov variation discussed by Unruh and others in [7]), such things would be inherently impossible. Our new analysis not only suggests that it *would* be possible; it also gives some guidelines for the engineering to actually go out and test it. But what if there are no such regions out there to find? Price argues very persuasively that Hawking's *first* version of the Big Bang model (the time-symmetric version) is far more plausible than the *revised* version in the second edition of Hawking's book [8], which essentially adds epicycles (unnecessary additional complexity) in order to match rigid dogmas about time-forwards causality. *If* the Big Bang model were true, I would consider Price's arguments totally applicable, and I would advocate going ahead and building the new instruments in order to look for these regions of space. But as it happens, I doubt that those regions are there. Maybe, but probably not – at least not for the ordinary kind of matter we know how to detect.

Could the time-asymmetry of the superweak effects itself be the result of a kind of spontaneous symmetry breaking, on the scale of hundreds of billions of years or more? Of course it could, in principle. But it will be enough of a challenge for us to start decoding what we can see within one hundred billion years, for the time being.

**Completing the Backwards-Time Interpretation of Quantum Mechanics**

In [1] and in our other paper in this issue, we proved how continuous field theories result in statistical dynamics close to or the same as quantum dynamics for bosonic field theories. This leaves open the question: what about mixed fermi/bose theories like quantum electrodynamics (QED) and the standard model of physics?

As this paper goes to press, I see (very) preliminary evidence that fermi/bose dynamics may emerge from the statistical dynamics of a *Lorentzian* system, a system of continuous fields (like $A_\mu$) and of *point particles*. For now, it appears that quantum behavior may result from the light and from quantum measurement, without any need to add DeBroglie's kind of "pilot wave" field. A model based on solitons [1] would be more satisfying, but we do not yet have *empirical* evidence for a nonzero radius for the electron, and QED does assume that the electron is a point particle (with infinite self-energy!). Some concepts of "the universe as a great mind" would actually fit better with a point particle model of the electron as a starting point. Still, as with QED itself and Lorentzian field theory, the field theory is well-defined only after we attach a renormalization procedure.

To evaluate this idea, it should be straightforward to use the same kind of mathematical analysis as in [1], adding fermionic operators to the system, *once we specify the "code."* By the "code," I mean that we must specify the equation for point particles corresponding to equation 6 of our other paper. For a single electron/positron at point **q**, momentum **p**, spin S and charge Q, the obvious code for a single-particle pure state would be:

$$\psi(\underline{x}) = \delta(\underline{x} - \underline{q}) \, \psi_0(\underline{p}, S, Q),$$

where $\psi_0$ could be the usual Dirac 4-spinor for (**p**, S, Q) for a free particle as given by Bjorken and Drell or Messiah, with the exp($\alpha$**x**) term removed. The multiparticle pure-state wave function (the equivalent of **w** in equation 6!) might be the usual product of individual wave functions, antisymmetrized in the usual way; the resulting undefined factor of (-1)/(1) disappears in the density matrix $\rho$. The usual relation between gradients and momentum operators reappear only after we focus on irreducible equilibrium ensembles, as discussed in section 1 of our other paper. If this approach works, it suggests that exact local "bosonization" of QED could occur only as the *limit* of a family of bosonic field theories which sustain solitons, in the limit as the soliton radius goes to zero.

This idea has some parallels to our older idea (quant-ph 008036) of bosonic wave functions representing statistical moments. As in [1], one might need to do some scaling and the like, and one might even need to consider alternative forms of $\psi_0$ or –in the worst case – terms like $(\nabla\delta)\psi_1$, to make the match to QED exact. But for now, there is hope that the code as described here could match QED exactly, in its predictions for bound states and scattering states.

**Views of the Physics of Life and Mind**

Prigogine tried to apply thermodynamics not only to the universe but to life as well.

But what happens when we give up on the idea of a *local* entropy function? We allow for the possibility of patterns and life. We allow for the possibility that we are *already* living in the maximum of entropy – but we need to be careful to understand what this means. Above all, we end up throwing out a nice, straightforward approximation that pretended to describe all dynamical systems, *without anything to take its place*. To develop a serious, thermodynamics-like mathematical theory of life, we would need to develop a whole new strand of mathematics. There are emerging communities in "quantitative systems biology" (QSB), who are groping towards the development of such new mathematics. Freeman, Kozma and I have discussed a few possibilities [9]. But it is all at an early stage, and no one knows as yet how much Mathematical unity and generality is really possible here. It is interesting to consider, however, how the elaborate and powerful chains of approximation methods developed for the neural network community [4] might be relevant to the inevitable nonlinear approximations needed here. Perhaps pattern recognition and pattern emergence do have relations with each other, just as statistical learning and numerical convergence do.

There are some physicists, mystics and parapsychologists who would be badly repelled by the "mechanical" and "materialistic" vision of the universe I portrayed above. Many physicists agree with Buddhism and with Idealism that the ultimate laws of the universe will turn out to be more like the laws of a Great Mind than the laws of a Great Machine. (Hundreds could be cited on this point.) I certainly do not claim to Know whether they are right or not. Certainly when I see movies like *What Dreams May Come* or *Forbidden Planet* or *The Matrix*, or read Greg Bear's *Moving Mars*, I feel that any of these *might* turn out to be true in the end. (Though The Matrix does have some questionable aspects.) Yet we can only make progress "one step at a time." The matrix condensation formalism for quantum measurement usually blamed on Copenhagen has nothing at all to do with the mathematics of Mind or Intelligence, as we are beginning to understand it. (Heisenberg's collaborator Duerr has been very emphatic that they should not be blamed for this particular ad-hoc add on to their more elegant operator formalisms, which are more like what I call "quantum dynamics.").

Many "general systems" people have sometimes gone overboard in attacking anyone as "reductionist" who believes that there exist mathematics laws underlying all of the universe. But there really do exist a few extreme reductionists in physics, who seem to believe that "explaining the mind" or "explaining psychic phenomena" is just a matter of identifying the relevant force fields and writing a wave equation. One can't even explain how a radio works using such a simple-minded approach (which does not include circuit analysis!), let alone a brain or a mind. To explain the physical basis of biological or mental phenomena, one needs to begin by considering the thermodynamic foundations – the greater flows of free energy, forwards *or backwards* in time, across the universe. But to understand how mental phenomena actually work, in any kind of practical or empirical sense, the mathematics of mind[2,4,10] are more directly applicable. It is interesting to consider what kind of life or mind might evolve in a larger ecology with more time-symmetric flows of energy, but I have not had time or motivation to really follow up on some crude ideas along those lines[1,6].

It is also interesting to note that physical energy (the Hamiltonian) can never be created or destroyed, but that "psychic energy" or "cathexis" as described by Freud is much closer to modern concepts of feedback [2,10] which actually tend to *increase* as an intelligent system grows in maturity. If people naively equate these two concepts, there is enormous opportunity for dangerous confusion. For example, many cultures have believed very deeply in "mana" or "charisma" as a kind of physical energy, subject to exact conservation; this has led many to practices like voluntary or involuntary human sacrifice, in which it was believed that the perpetrator must acquire new energy equal to that of the person being murdered. Absurd as such ideas are, they had a powerful effect in stifling the progress (and energy) of many cultures. It would be ironic if a misuse of modern mathematical concepts led to a kind of re-emergence of such destructive confusion. There are times when the treatment of graduate students or of minority opinions in science, and the assumption of a "zerosum game" in many walks of life, reminds me of those sad old days.

**Conclusion**

It is sobering to think that we might be almost totally ignorant of the vast if dispersed sources of free energy which underlie our very existence. We may have more in common than we think with the medieval peasants, who could see the stars whirling in the sky but could not begin to figure out the connection between those stars and the physics of their everyday life. Like them, we may be doomed to essential ignorance in our lifetimes. (Many medieval people tried to *imagine* connections between the stars and their lives, but the results were quite embarrassing.) But if we develop the mathematical prerequisites and work hard and patiently and boldly to extend our real understanding, then perhaps someday our descendants will be able to attain a level of life that we peasants can hardly imagine. Alternatively, of course, the option of stagnation, fragmentation and extinction is also available to all species in the greater biosphere.